\documentclass[a4paper,twocolumn,english,aps,prl,reprint,superscriptaddress,showpacs,showkeys,longbibliography, nofootinbib]{revtex4-2}
\usepackage[utf8]{inputenc}
\usepackage{babel}
\usepackage{mathtools}
\usepackage{appendix}
\usepackage{amsmath}
\usepackage{amssymb}
\usepackage{graphicx}
\usepackage{float}
\usepackage{physics}
\usepackage[colorlinks=true, linkcolor=red, citecolor=blue, urlcolor=blue]{hyperref}
\usepackage[usenames,dvipsnames]{color}
\usepackage{bm}
\usepackage{bbold}
\usepackage{braket}
\usepackage{comment}
\usepackage{dcolumn,extarrows}
\usepackage{soul}
\usepackage{footmisc}
\usepackage{pdfpages}
\usepackage{pgf,tikz}
\usepackage{times}
\usepackage{subcaption}
\usepackage[normalem]{ulem}

\allowdisplaybreaks[2]

\definecolor{mygreen}{rgb}{0,0.5,0}
\definecolor{myblue}{rgb}{0,0,0.75}
\definecolor{mymagenta}{cmyk}{0,1,0,0.12}

\newcommand{\ud}{\mathrm{d}}

\newcommand{\minus}{
	\setbox0=\hbox{-}
	\vcenter{
		\hrule width\wd0 height \the\fontdimen8\textfont3
	}%
}

\newcommand{\be}{\begin{equation}}\newcommand{\ee}{\end{equation}}

\definecolor{mygreen}{rgb}{0,0.5,0}\definecolor{myblue}{rgb}{0,0,0.75}\definecolor{mymagenta}{cmyk}{0,1,0,0.12}

\makeatletter
\AtBeginDocument{\let\LS@rot\@undefined}
\makeatother

\begin{document}

\title{A post-Newtonian Gravitational Collapse Model from Linearized Gravity}

\author{Rudi B. P. Pietsch}\email[]{rudi.pietsch@uni-ulm.de}
\affiliation{Institut f\"ur Theoretische Physik, Albert-Einstein-Allee 11, Universit\"at Ulm, 89069 Ulm, Germany}
\affiliation{Center for Integrated Quantum Science and Technology (IQST), 89081 Ulm, Germany}
\author{Luciano Petruzziello}\email[]{luciano.petruzziello@uni-ulm.de}
\affiliation{Institut f\"ur Theoretische Physik, Albert-Einstein-Allee 11, Universit\"at Ulm, 89069 Ulm, Germany}
\affiliation{Center for Integrated Quantum Science and Technology (IQST), 89081 Ulm, Germany}
\author{Martin B. Plenio}\email[]{martin.plenio@uni-ulm.de} \affiliation{Institut f\"ur Theoretische Physik, Albert-Einstein-Allee 11, Universit\"at Ulm, 89069 Ulm, Germany}
\affiliation{Center for Integrated Quantum Science and Technology (IQST), 89081 Ulm, Germany}
 
\begin{abstract}
We introduce a general gravity-related collapse mechanism based on linearized gravity. Starting from the weak-field limit of general relativity, gravitoelectromagnetism suggests an effective coupling between the gravitoelectric potential and the mass density distribution. At the same time, it provides a similar relation for the gravitomagnetic vector potential and the mass current. Following a hybrid (classical-quantum) dynamics approach, these couplings lead to a master equation whose non-unitary part is determined by the underlying mass distribution and currents. When the gravitoelectric potential coupling is considered, the well-known Di\'osi-Penrose collapse model acting on positional degrees of freedom is recovered. However, upon including the gravitomagnetic vector potential, additional collapse mechanisms emerge for rotational degrees of freedom as well as for mixed mass-rotation contributions. 
\end{abstract}
	
\date{\today}
\maketitle
	
{\emph{Introduction --}} Collapse models provide a phenomenological tool to parametrize possible deviations from quantum mechanics at the macroscopic scale while preserving the standard unitary dynamics for microscopic systems. Along this line, several mechanisms have been proposed, including the spontaneous localization model of Ghirardi, Rimini and Weber~\cite{GRW1986,Pearle1989,Ghirardi1990CSL}. Further developments include refined parameter choices and variants of continuous spontaneous localization tailored to different physical conditions~\cite{Adler2007} as well as systematic discussions of the theoretical structure and experimental bounds~\cite{BassiEtAl2013RMP,BassiUlbricht2014}. Within this framework, the Di\'osi--Penrose (DP) model~\cite{Diosi1987,Diosi1989,Penrose1996, diosi1995quantum} represents one of the most important collapse mechanism whose motivation can be traced back to gravity in its Newtonian limit through the mass density distribution. Although the parameter-free version of the DP model has been ruled out experimentally~\cite{Donadi2020NatPhys}, its form with a free variable encoding the spatial resolution of the mass density~\cite{Ghirardi1989,Diosi2013} still remains a possible candidate to describe the wave function collapse.

Since the DP mechanism is formulated in terms of the mass density operator, it is most naturally associated with positional decoherence. Nevertheless, for a rigid body with a mass distribution lacking symmetry, different orientations correspond to different spatial mass densities. In such cases, the non-unitary part of the DP dynamics can generate orientation-dependent decoherence. However, this effect is entirely governed by the anisotropy of the mass distribution. Indeed, it vanishes identically for any body that is either spherically symmetric or rotating about a symmetry axis, since in both cases the rotation leaves the mass density invariant (see Appendix \ref{app:different_predictions}). More broadly, the DP framework offers no systematic access to effects driven by mass currents, nor does it predict decoherence in the angular momentum basis independently of the body's geometry. Although semiclassical and hybrid gravity frameworks dealing with rotational degrees of freedom have been proposed~\cite{oppenheim2023postquantum, hu2008stochastic, kuo1993semiclassical}, a systematic treatment of the resulting collapse and decoherence mechanisms is still lacking. This issue is becoming increasingly relevant, as a number of experimental platforms are now able to perform accurate trapping of the center-of-mass motion while at the same time enabling coherent control of the orientation. 

Concrete examples along these lines include torsional optomechanics with optically levitated nanoparticles in vacuum~\cite{Hoang2016Torsional}, as well as levitated nanodiamonds hosting controllable NV-center spins~\cite{Hoang2016NV}. In these settings, translational decoherence may be difficult to isolate experimentally, since the center-of-mass motion is already tightly confined. On the other hand, the rotational dynamics offers an independent observable channel that can provide a distinctive signature of fundamental decoherence.

Indeed, rotational degrees of freedom may also play a crucial role in proposals for gravity-mediated entanglement. In this context,  Refs.~\cite{LantanoPetruzziello2024,PetruzzielloLantano2025, hjfx-rlfj} discuss how frame {dragging, a post-Newtonian effect, may} be used to witness entanglement through angular momentum correlations, whereas Ref.~\cite{Higgins2024} proposes to employ the rotational energy to enhance entanglement growth between spatially delocalized probes interacting via Newtonian gravity. In such scenarios, a collapse mechanism acting on angular momentum would introduce additional decoherence channels, thereby imposing possible limitations, but would also offer a new target for experimental bounds.

In this Letter, we introduce a gravitationally-inspired collapse model which can be viewed as a post-Newtonian extension of the Di\'osi--Penrose proposal. Our starting point is the weak-field, slow-motion expansion of general relativity, where gravitoelectromagnetism~\cite{heaviside} reveals a formal analogy between the electromagnetic and the gravitational interaction, separating the dynamics into a gravitoelectric sector governed by the scalar potential and a gravitomagnetic sector governed by the vector potential. Building on this analogy and following the classical-quantum hybrid dynamics strategy of Di\'osi~\cite{Diosi2011Hybrid}, we derive a generalized master equation whose non-unitary part comprises three classes of decoherence channels: (i) a purely translational channel that reproduces the standard DP double-commutator in the mass density, (ii) a purely rotational channel quadratic in the mass current, whose structure is determined by the current distribution and is closely related to the gravitomagnetic sector of linearized gravity (\emph{i.e.}, the one responsible for frame dragging), and (iii) mixed channels in which mass density and mass current contribute jointly. For rigid rotors, the mass current can be expressed in terms of the angular momentum, thus making rotational decoherence a direct corollary of the more general post-Newtonian dynamics. 

{\emph{Hints from gravitoelectromagnetism --}}\label{sec2} The Di\'osi--Penrose model~\cite{Diosi1987,Diosi1989,Penrose1996, diosi1995quantum} finds its motivation in the Newtonian limit of general relativity. However, in the weak-field regime, such limit is not the only relevant contribution. In particular, when the velocity of matter cannot be neglected with respect to the speed of light, additional terms arise from the Einstein field equations and are sourced by the mixed time--space components of the stress-energy tensor.

To make this statement explicit, we start from the field equations when small perturbations around the flat spacetime are considered, \emph{i.e.}, when $g_{\mu\nu}=\eta_{\mu\nu}+h_{\mu\nu}$. Upon introducing $\bar{h}_{\mu\nu}=h_{\mu\nu}-\eta_{\mu\nu}h/2$ (with $h=\eta_{\mu\nu}h^{\mu\nu}$) and imposing the harmonic gauge condition $\partial_\mu\bar{h}^{\mu\nu}=0$ (Einstein's summation convention is understood), the metric perturbations are linked to the stress-energy tensor via the wave equation
\begin{equation}\label{feq}
    \Box\bar{h}^{\mu\nu}=-\frac{16\pi G}{c^4}\,T^{\mu\nu}\,.
\end{equation}
When addressing the Newtonian limit, the only non-vanishing term of $T^{\mu\nu}$ is $T^{tt}=m c^2$, with $m$ being the mass density. In so doing, the condition $v^k\ll c$ is already implicitly enforced, since the off-diagonal terms $T^{tk}=J^kc=m v^k c$ are neglected.

By relaxing this condition and defining the four-vector potential components $A^\mu=(\phi/c,A^k)$ via metric perturbations as $\bar{h}^{tt}=4\phi/c^2$ and $\bar{h}^{tk}=4A^k/c$, the field equations \eqref{feq} can be rewritten as 
\begin{equation}\label{gem}
    \Box A^\mu=-\frac{4\pi G}{c^2}\,J^\mu\,,
\end{equation}
with $J^\mu=(m c,J^k)$ and the harmonic condition becoming $\partial_\mu A^\mu=0$. This rephrasing of Einstein's field equations is at the foundation of gravitoelectromagnetism, which shows the formal analogy between electromagnetic and gravitational phenomena in the weak-field limit. 

The Newtonian regime is recovered in the static limit of Eq. \eqref{gem}, for which $\mu=t$ returns the Poisson equation $\nabla^2\phi=4\pi G m$.
However, the same set of equations also implies the existence of $\nabla^2 \vec{A} = 4 \pi G \vec{J}/c^2$, which, depending on the considered approximation, might not be negligible.

In view of a potential application of the above formalism in the context of collapse models mimicking the DP scheme, one can argue that the equation for $\vec A$ is relevant for the angular momentum degrees of freedom. To see why this is the case, consider a rigid rotor free to spin around one of its symmetry axes. If the rigid body is trapped sufficiently tight, 
its center-of-mass degrees of freedom are already well localized, thereby implying the absence of a positional collapse mechanism.

Following the strategy of Di\'{o}si~\cite{Diosi2011Hybrid}, one can then consider a hybrid classical--quantum dynamics where the classical gravitomagnetic potential couples canonically to the quantum mass current through an interaction term $\int \ud^3x\,\hat{\vec{J}}(\vec{x}) \cdot \vec{A}(\vec{x})$. The coupling to the classical degrees of freedom yields a non-unitary evolution for the quantum state, leading to a master equation whose decoherence part  depends on the quantized current $\hat{J}^k$. We note that the scalar sector already generates rotational decoherence for bodies whose mass distribution is not spherically symmetric, since distinct orientations correspond to distinct spatial densities (see Appendix~\ref{app:different_predictions}). The gravitomagnetic sector identified here is complementary, as it acts on the mass current rather than the mass density and is therefore present even for symmetric bodies.

In particular, for a rigid rotor, the current distribution $\vec{J}$ can be rewritten in terms of the angular momentum operator $\vec{L}$. Indeed, since for a rotation around a symmetry axis one has $\vec{L}=I\vec{\omega}$,
the current can be cast as $\vec{J}=m\,\vec{\omega}\wedge\vec{x}=m\,\vec{L}\wedge\vec{x}/I$. As a consequence, the induced decoherence mechanism acts on the rotational basis.

This picture is further corroborated by the classical far-field structure of the gravitomagnetic sector. Indeed, in the weak-field approximation, a localized rotating source generates a gravitomagnetic field whose leading-order contribution has the same tensorial form as the magnetic field of a dipole, with the role of the dipole moment being assumed by the angular momentum. Explicitly, for $r$ much larger than the source size, one finds~\cite{ciufolini}
\begin{equation}\label{eq:magneticfield}
    \vec{B}(\vec{x})=\vec \nabla\wedge\vec{A}(\vec{x})\simeq \frac{2G}{c^2 |\vec{x}|^3}\left[\vec{L}-3\,\frac{\big(\vec{L}\cdot{\vec{x}}\big){\vec{x}}}{|\vec{x}|^2}\right] \, .
\end{equation} 
Hence, the gravitomagnetic field directly encodes the angular momentum of the source, thereby providing a natural motivation for gravitationally-induced collapse terms acting in the rotational basis.

{\emph{Classical-quantum master equation --}}\label{sec3} We now derive the post-Newtonian-inspired hybrid master equation using Di\'{o}si's approach~\cite{Diosi2011Hybrid} (more details are provided in Appendix \ref{app:derivation}). The substantial difference from the formalism introduced in Ref.~\cite{Diosi2011Hybrid} is that, in what follows, we will rely on the quantized post-Newtonian mass-energy current $\hat{J}^{\mu} = (\hat{m}c, \hat{J}^{k})$ rather than considering only the mass density $\hat{m}$ as the quantum degree of freedom. Similarly, the classical degrees of freedom are given by the four-vector potential components $A^{\mu} = (\phi/c, A^k)$, with $\vec{A} = (A^1, A^2, A^3)$ denoting the gravitomagnetic vector potential, instead of having only $\phi$. 

Henceforth, we assume Hamiltonians and states, whether quantum or classical, to be conditioned on classical phase space variables denoted by $\vec{z}$~\cite{barchielli2024hybrid, dammeier2023quantum, barchielli2023markovian}. However, instead of only including the gravitational potential and its conjugate momentum (as in \cite{oppenheim2023postquantum}), our extended phase space variables consist of $A^{\mu}$ and its conjugate momentum $\pi^{\mu}$, reading $ \vec{z} = \{ A^{\mu}(\vec{x}), \pi^{\mu}(\vec{x}) \} $. All values of the phase space components are determined at spatial configurations $\vec{x} \in \mathbb{R}^3$. 

With this knowledge, we define the quantum state and the Hamiltonian as $\hat{\tau}(\vec{z},t)$ and $\hat{H}_Q(\vec{z},t)$, respectively. Similarly, we can introduce a classical Hamiltonian $H_C(\vec{z},t)$ and state $p(\vec{z},t)$. In classical-quantum hybrid treatments, one combines both states into a separable, hybrid state $ \hat{\rho}(\vec{z},t) = \hat{\tau}(\vec{z},t) p(\vec{z},t) $ while allowing the two systems to interact, which leads to the appearance of an interaction Hamiltonian $\hat{H}_{I}(\vec{z},t)$ in the hybrid equation of motion such that the total hybrid Hamiltonian reads $\hat{H}(\vec{z},t) = \hat{H}_Q(\vec{z},t) + H_C(\vec{z},t) + \hat{H}_{I}(\vec{z},t)$. Now, a noiseless evolution of the hybrid state takes the form
\begin{widetext}
\begin{equation}\label{aleks}
\partial_t \hat{\rho}(\vec{z},t)  = - \frac{i}{\hbar} [\hat{H}(\vec{z},t), \hat{\rho}(\vec{z},t) ] + \frac{1}{2} \Big[ \{ \hat{H}(\vec{z},t), \hat{\rho}(\vec{z},t) \}_P   - \{ \hat{\rho}(\vec{z}, t), \hat{H}(\vec{z},t) \}_P  \Big] \, ,
\end{equation}
\end{widetext}
where $\{f,g \}_P := (\partial_{A_{\mu}}f) (\partial_{\pi^{\mu}}g) - (\partial_{\pi_{\mu}}f) (\partial_{A^{\mu}}g)  $ denotes the Poisson bracket. This classical-quantum evolution is called Aleksandrov bracket and was first proposed in Ref.~\cite{Aleksandrov+1981+902+908}. One problem associated with this equation is that it fails to ensure the positivity of the hybrid state, meaning that refinements to the interaction between classical and quantum systems are required. Furthermore, any hybrid theory that lacks fundamental noise is pathological since it violates quantum uncertainty relations (see Appendix~\ref{app:noise_necessity}). 

To account for that, one has to start from the interaction term coupling the classical four-vector potential to the quantized mass current which originates from the interaction
$\mathcal{L}_{\mathrm{int}} = -h_{\mu\nu}\hat T^{\mu\nu}/2$.
In our conventions, the $tt$-component of the coupling gives
$\hat{m}\phi$, whereas the $tk$-components yields
$4\hat{\vec{J}}\cdot\vec{A}$.  To write both
terms compactly through a single four-vector contraction, we define
$\tilde A^{\mu}=(\phi/c,\,4A^k)$, so that the
interaction Hamiltonian reads
\begin{align}\label{eq:interaction_definition}
\hat{H}_{I}(\vec{z}, t) = \int \mathrm{d}^3x \ \hat{J}^{\mu}(\vec{x}) \tilde A_{\mu}(\vec{x}) \, ,
\end{align}
with $\hat{J}^{\mu}(\vec{x}) \tilde A_{\mu}(\vec{x}) = \hat{m}(\vec{x})\phi(\vec{x}) + \hat{\vec{J}}(\vec{x}) \cdot \vec{\tilde A}(\vec{x})$. To ensure positivity and preservation of the uncertainty relations, the above Hamiltonian must contain a classical noise term for both $\tilde A^{\mu}(\vec{x})$ and $\hat{J}^{\mu}(\vec{x})$. This noise, denoted as $\mathcal{A}^{\mu}(\vec{x}, t)$ and $\mathcal{J}^{\mu}(\vec{x}, t)$, enters the interaction Hamiltonian as
\begin{equation}
    \hat{\mathcal{H}}_{I}(\vec{z},t) = \int \mathrm{d}^3 x \ \Big[ \hat{J}^{\mu}(\vec{x}) + \mathcal{J}^{\mu}(\vec{x}, t) \Big] \Big[ \tilde A_{\mu}(\vec{x}) + \mathcal{A}_{\mu}(\vec{x}, t) \Big] \, . 
\end{equation}
A general noise Hamiltonian modifies the total Hamiltonian, which then becomes $\hat{\mathcal{H}}(\vec{z}, t) = \hat{H}_Q(\vec{z}, t) + H_C(\vec{z},t) + \hat{\mathcal{H}}_{I}(\vec{z},t)$. As customary for collapse models, the noise is taken to be Gaussian and white, so that their first moments (\emph{i.e.}, their respective noise averages) vanish as $\langle \mathcal{J}^{\mu}(\vec{x}, t) \rangle = \langle \mathcal{A}^{\mu}(\vec{x}, t) \rangle = 0$, whereas the second moments are spatially dependent coefficients sharply peaked in time: 
\begin{align}
        & \langle \mathcal{J}^{\mu}(\vec{x},t) \mathcal{J}^{\nu}(\vec{y},t') \rangle = D^{\mu \nu}_{\mathcal{J}}(\vec{x},\vec{y}) \delta(t-t') \, , \notag \\[2mm] & \langle \mathcal{A}^{\mu}(\vec{x},t) \mathcal{A}^{\nu}(\vec{y},t') \rangle = D^{\mu \nu}_{\mathcal{A}}(\vec{x},\vec{y}) \delta(t-t') \, .
\end{align}
The noise must adhere to $\partial_{t} \mathcal{A}^{t} \ + \ \vec \nabla \cdot \mathcal{\vec A}/4 = 0$ to avoid harmonic gauge violations for $\Tilde{A}^{\mu}$. Albeit technical, this condition might become vital, \emph{e.g.}, when trying to determine a functional form of $D^{\mu \nu}_{\mathcal{A}}(\vec{x}, \vec{y})$ \cite{future_work}. Additionally, the white noise introduced here leads to an unbounded energy growth, which is an acknowledged problem in hybrid theories of gravity \cite{oppenheim2023objective}, implying the need to find suitable friction terms that satisfy a hybrid fluctuation-dissipation relation for the dynamical equations.

Now, to ensure the positivity of the hybrid state in~\eqref{aleks}, one can refine the dynamics of the hybrid state by averaging over the noise, thereby remaining with the noise-averaged Aleksandrov bracket~\eqref{aleks} with $\mathcal{H}(\vec{z},t)$ instead of $H(\vec{z},t)$
. In accordance with Refs. \cite{diosi1995quantum, Diosi2011Hybrid}, we note that the positivity of the resulting dynamics is ensured by $D_{\mathcal{A}} D_{\mathcal{J}} \succeq \hbar^2 \mathbb{1}/4$, where $D_{\mathcal{A/J}}$ are $4\times 4$ matrices made up of the $D^{\mu \nu}_{\mathcal{A/J}}$ components, $\mathbb{1}$ is the $4\times 4$ identity matrix, and $\succeq$ denotes positive semi-definite ordering. The inequality is to be understood after smearing the spatially dependent kernels against suitable test functions so that the resulting finite-dimensional matrices satisfy this ordering (for more details, see Refs. \cite{oppenheim2023postquantum, oppenheim2023gravitationally}). By assuming $\langle \mathcal{J}^{\mu}(\vec{x},t)\mathcal{A}^{\nu}(\vec{y},t') \rangle = 0$ and expanding all noise terms on the classical and quantum variables to second order in $\hat{J}^{\mu}$ as well as $\tilde A^{\mu}$ (see Appendix \ref{app:derivation}), the averaged evolution of a hybrid state reads
\begin{widetext}
    \begin{align}\label{eq:rotational_dynamics}
    \partial_t \hat{\rho}(\vec{z},t) & = - \frac{i}{\hbar} [\hat{H}(\vec{z},t), \hat{\rho}(\vec{z},t) ] + \frac{1}{2} \Big[ \{ \hat{H}(\vec{z},t), \hat{\rho}(\vec{z},t) \}_P - \{ \hat{\rho}(\vec{z}, t), \hat{H}(\vec{z},t) \}_P  \Big] \notag \\[2mm] & - \frac{1}{2\hbar^2} \int \mathrm{d}^3 x \ \mathrm{d}^3y \ D^{\mu \nu}_{\mathcal{A}}(\vec{x},\vec{y}) \left[ \hat{J}_{\mu}(\vec{x}),\left[ \hat{J}_{\nu}(\vec{y}) , \hat{\rho}(\vec{z}, t)\right] \right] + \frac{1}{2} \int \mathrm{d}^3 x \ \mathrm{d^3}y \ D^{\mu \nu}_{\mathcal{J}}(\vec{x},\vec{y}) \left\{ \tilde A_{\mu}(\vec{x}), \left\{ \tilde A_{\nu}(\vec{y}), \hat{\rho}(\vec{z},t) \right\}_P \right\}_P \, .
\end{align}
\end{widetext}

{\emph{Implications --}}\label{sec4} We shall now shed light on some consequences of the generalized post-Newtonian hybrid dynamics described by~\eqref{eq:rotational_dynamics}. First, the DP dynamics of Ref. \cite{Diosi2011Hybrid} is recovered for systems whose rotation is negligible, that is, when all terms that scale with $\hat{\vec{J}}(\vec{x})$ and $\vec{A}(\vec{x})$ vanish, meaning that only $\hat{J}^t(\vec{x}) = \hat{m}(\vec{x}) c$ and $A^t(\vec{x}) = \phi(\vec{x})/c $ contribute to the noise interaction. The equivalence is ensured when we demand that $c^2 D^{tt}_{\mathcal{A}}(\vec{x}, \vec{y})$ corresponds to the decoherence rate of Ref. \cite{Diosi2011Hybrid}.

Next, there is a class of terms that are quadratic in the quantized mass current $\hat{\vec{J}}(\vec{x})$ and the gravitomagnetic potential $\vec{A}(\vec{x})$. In particular, the non-unitary terms after tracing out the classical degrees of freedom (which makes the last term of \eqref{eq:rotational_dynamics} vanish as shown in Appendix \ref{app:reduced_dynamics}) are given by
\begin{equation}
    \mathcal{L}_{\text{rot}} \hat{\sigma}(t) = -\frac{1}{2\hbar^2} \int \mathrm{d}^3 x \mathrm{d}^3 y D^{kl}_{\mathcal{A}}(\vec{x}, \vec{y}) \left[\hat{J}_k(\vec{x}) , \left[ \hat{J}_l(\vec{y}), \hat{\sigma}(t) \right] \right] ,
\end{equation}
where $k,l \in \{1,2,3\}$ and $\hat{\sigma}(t) := \int \mathrm{d}^8 z \ \hat{\rho}(\vec{z},t)$ denotes the reduced quantum state. 

For a rigid rotor, one can then write $\hat{\vec{J}}(\vec{x}\,) = m(\vec{x}\,)\,\hat{\vec{L}} \wedge \vec{x}/I$, and the rotational decoherence term $\mathcal{L}_{\text{rot}} \hat{\sigma}(t)$ takes the form
\begin{equation}\label{eq:rotdecoh}
     - \frac{1}{2\hbar^2} \int \mathrm{d}^3 x \ \mathrm{d}^3 y \ D^{kl}_{\mathcal{A}}(\vec{x}, \vec{y})\chi_{ki;lm}(\vec x,\vec y)\left[\hat{L}^i, \left[ \hat{L}^m, \hat{\sigma}(t) \right] \right] \, ,
\end{equation}
with $\chi_{ki;lm}(\vec x,\vec y)=m(\vec x)m(\vec y) \varepsilon_{kij}\varepsilon_{lmn}x^j y^n/I^2$ and $\varepsilon_{ijk}$ being the Levi-Civita symbol.

From Eq.~\eqref{eq:rotdecoh}, it is clear that these terms induce decoherence in the angular momentum
basis.  More precisely, because the three components $\hat{L}^k$ do not 
mutually commute, the double-commutator structure does not select a
single preferred eigenbasis but rather generates a decoherent process on
the rotation group. For instance, for a rotor constrained to spin about a fixed symmetry axis $z$, the resulting dephasing appears in the $(\hat L^x,\hat{L}^y)$ basis (see Appendix \ref{app:different_predictions}) {and, notably, preserves $\hat{L}_z$}. Given that fast-rotating systems are becoming experimentally controllable \cite{PhysRevLett.121.033602, PhysRevLett.121.033603, ahn2020ultrasensitive,6ghz}, we expect that constraints on the decoherence rates can be accessible in quantum optical laboratory setups. 

Finally, one can identify the contributions, including $D^{\mu\nu}$ with $ (\mu = t \wedge \nu \neq t) \vee (\mu \neq t \wedge \nu = t)$ as non-unitary factors where both positional and rotational degrees of freedom contribute linearly. In the equation of motion for the reduced quantum system, one can then observe the emergence of non-unitary terms
\begin{equation}
    - \frac{c}{2\hbar^2} \int \mathrm{d}^3 x \ \mathrm{d}^3 y \ D^{tl}_{\mathcal{A}}(\vec{x}, \vec{y}) \left[\hat{m}(\vec{x}), \left[\hat{J}_l(\vec{y}) , \hat{\sigma}(t) \right] \right] \, , 
\end{equation}
\begin{equation}
    - \frac{c}{2\hbar^2} \int \mathrm{d}^3 x \ \mathrm{d}^3 y \ D^{kt}_{\mathcal{A}}(\vec{x}, \vec{y}) \left[\hat{J}_k(\vec{x}), \left[\hat{m}(\vec{y}) , \hat{\sigma}(t) \right] \right] \, .
\end{equation}
These contributions become increasingly relevant for systems that are both massive and rotate fast. Such objects can be found in cosmology in the form of pulsars, which show astonishingly low fluctuations in their rotation \cite{lorimer2008binary, sun2023estimating, shabanova2013timing}, making them promising probes for finding constraints on the decoherence rate where rotational degrees of freedom interact with the center-of-mass motion.

To quantify the relative weight of the three decoherence channels,
we note that they scale as $m^2 c^2$, $J^2 \sim
m^2 v_{\mathrm{surf}}^2$ and $m J c \sim m^2 v_{\mathrm{surf}} c$, where $v_{\mathrm{surf}} = \Omega R$ is the surface
velocity of the rotor.  For the $6$ GHz silica nanosphere of
Ref.~\cite{6ghz} (with $R \simeq 85$ nm and $\Omega \simeq 3.8 \times
10^{10}$ rad $\text{s}^{-1}$), one finds $v_{\mathrm{surf}}/c \simeq 10^{-5}$,
so that, for comparable noise kernels, the purely rotational and mixed
channels are suppressed by $\sim 10^{-10}$ and $\sim 10^{-5}$
relative to the DP term for non-symmetric rotations.  However, the noise kernels $D^{kl}_{\mathcal{A}}$ and
$D^{tk}_{\mathcal{A}}$ are independent free parameters of the
model, entirely unconstrained by existing translational
experiments that only probe $D^{tt}_{\mathcal{A}}$.  Therefore, fast-rotor
experiments have access to a distinct sector of the theory,
making them complementary to standard DP tests even when the
source products are kinematically suppressed.  The situation
becomes even more compelling for millisecond pulsars (with $R \sim
10$ km and $\nu \sim 700$ Hz, as in Ref.~\cite{Hessels2006}), where $v_{\mathrm{surf}}/c \simeq
0.15$ and all three channels contribute on a comparable footing.

Since the above argument indicates that pulsars are promising probes to investigate this effect, one might expect significant rotational decoherence rates when dealing with these astrophysical objects. Specifically, considering the dipolar interaction arising from the gravitomagnetic field (the quantized version of Eq. \eqref{eq:magneticfield}), the interaction Hamiltonian of two pulsars rotating close to each other scales as $\hat{H}_I \sim G \hat{L}^2 / c^2 R^3 $. As classical-quantum hybrid theories do not allow for entanglement buildup via gravity, the rotational noise must counter an entangling rate $\Gamma \sim GL^2 / \hbar c^2 R^3$ (see Ref.~\cite{LantanoPetruzziello2024} for more details), such that $\mathcal{L}_{\text{rot}} \hat{\sigma}(t) \gtrsim \Gamma \hat{\sigma}(t)$ when we heuristically restrict ourselves to the rotation channel. If the pulsars are rotating about their symmetry axes $L_z$, we have $L \sim 2\pi I_z \nu$ with $I_z \sim 2 M R^2/5$ for uniform spheres (taking $M=2M_{\odot}$ ~\cite{Hessels2006}), yielding an average estimate on the rotational decoherence of $\mathcal{L}_{\text{rot}} \hat{\sigma}(t) \gtrsim 64\pi^2 G M_{\odot}^2 R \nu^2 \hat{\sigma}(t) /25\hbar c^2 \sim 10^{78} \hat{\sigma}(t)$ s$^{-1}$, which implies a considerable loss of rotational coherence. However, this effect is not directly observable in the pulsars' frequency, as $\langle \hat{L}_z \rangle$ remains invariant under such rotations (see Appendix \ref{app:different_predictions}). Rather, distinct rotational pulsar observables, most notably precession \cite{stairs2000evidence}, would be affected. Rotating binary double pulsar systems (such as PSR J0737–3039 \cite{burgay2003increased, lyne2004double}) would constitute intriguing objects to study, since their lack of rotational symmetry translates to decoherence effects in observable pulsar timing predictions and may serve as a probe of testing DP effects against the model presented here.

{Moreover, to infer whether a rate of $10^{78}$ s$^{-1}$ is reasonable in the context of pulsars, let us compare it to the DP decoherence rate $\Gamma_{\text{DP}} = E_{\Delta}/\hbar$ \cite{Penrose1996, diosi2022conjectured} in a similar setting, with $E_{\Delta}$ being the gravitational self-energy of the difference between two spatially superposed states. If $d$ denotes the superposition distance, the self-energy scales as $E_{\Delta} \sim G M^2 d^2/R^3$, leading to $\Gamma_{\text{DP}} \sim 10^{73} \ d^2$ s$^{-1}$  m$^{-2}$. Hence, a DP-based estimation of a pulsar being in superposition leads to a similar decoherence rate as the rotational one when assuming $d \sim 10^{2.5} \ \text m$.}

{\emph{Conclusion --}}\label{sec5} We have introduced a post-Newtonian hybrid description of gravitational collapse within linearized gravity, extending the standard DP construction by including the gravitomagnetic sector. The resulting equation of motion~\eqref{eq:rotational_dynamics} contains: (i) the usual position-dependent decoherence channel as a limit in which rotational effects are negligible, (ii) additional non-unitary contributions quadratic in the mass current, which for rigid rotors reduce to a decoherence mechanism in the rotational basis, and (iii) mixed terms that couple center-of-mass and rotational degrees of freedom linearly.

Besides providing a unified language for translational and rotational decoherence, this framework suggests concrete experimental settings to constrain the corresponding noise kernels, ranging from levitated fast rotors to ultra-stable pulsar timing.

Equipped with this formalism, there are several directions that appear natural. First, it would be valuable to derive explicit parametrizations of the correlators $D^{\mu\nu}_{\mathcal{A}}$ and $D^{\mu\nu}_{\mathcal{J}}$ to identify which subclasses reproduce known collapse models and which instead predict additional rotational signatures. Second, the present formalism can be specialized to mesoscopic devices, where both translation and rotation are coherently controlled, allowing one to compute rotational decoherence rates, cross-couplings and the ensuing bounds from existing~\cite{pedernales2020decoherence,6ghz} and near-term experiments. Finally, given the appearance of mixed mass--current contributions, compact astrophysical rotors (pulsars) offer a situation in which the long-time stability of spin and orbital motion~\cite{lorimer2008binary, sun2023estimating, shabanova2013timing} could be used to bound the parameters entering~\eqref{eq:rotational_dynamics}. 

\bigskip
\noindent
\textbf{Acknowledgements:} This work was supported by the ERC Synergy grant HyperQ (grant no. 856432). L. P. acknowledges networking support by the COST Actions CA23130 (Bridging high and low energies in search of quantum gravity), CA23115 (Relativistic Quantum Information) and CA24101 (Testing Fundamental Physics with Seismology). We thank Tobias Haas for comments that helped improve the manuscript.

\bibliography{bib}

\clearpage

\onecolumngrid
\appendix
\section{Predictions Differing from the DP Model}\label{app:different_predictions}
The dynamics considered here contributes to the landscape of collapse models by serving as an extension beyond a mass density description. However, proper care is required to distinguish the unique rotational decoherence found here from effects that may also be accessed by previous collapse model treatments. One apparent argument is that rotation can never cause decoherence in a DP scheme as long as it does not alter the underlying mass density of the object. Hence, for every rotationally invariant body and likewise for every object rotating about an axis of symmetry, the dynamics presented here predicts decoherence, whilst conventional collapse model decoherence rates vanish. Since this statement represents a crucial aspect of our theory, let us now devote some effort into establishing it mathematically for a certain class of situations (taking inspiration from the decoherence treatment found in \cite{oppenheim2023gravitationally}).

Imagine a rigid rotor in superposition of two orientation states $|R_1 \rangle$ and $|R_2 \rangle$. Moreover, assume both rotational configurations to be connected via $|R_2 \rangle = \hat{\mathcal{R}}(\vec{\theta}) |R_1 \rangle$, with $\hat{\mathcal{R}}(\vec{\theta})$ being the rotation operator denoting a rotation by angle $\theta = |\vec{\theta}| \ \in [0, 2\pi)$ about the axis $\vec{n} = \vec{\theta}/\theta$:
\begin{equation}
    \hat{\mathcal{R}}(\vec{\theta}) = e^{-\frac{i}{\hbar} \vec{\theta} \cdot \hat{\vec{L}}} = e^{-\frac{i}{\hbar} \theta \vec{n} \cdot \hat{\vec{L}}} \, .
\end{equation}
Furthermore, we consider the angular separation to be sufficiently large, such that the rotational states can be effectively treated as an orthonormal basis since their overlap is negligible, that is $\langle R_i | R_j \rangle \approx \delta_{ij} \ \forall i,j \in \{1,2\}$, with $\delta_{ij}$ denoting the Kronecker delta. Note that, in principle, the following calculations can be formulated in a continuous setting by replacing sums over discrete states with integrals and the Kronecker delta with the Dirac delta. In the scenario considered here, the reduced quantum state can be written as
\begin{equation}\label{eq:superposition_state}
    \hat{\sigma}(t) = \sigma^{ij}(t) |R_i \rangle \langle R_j| \, .
\end{equation}
The off-diagonal elements $i\neq j$ represent the system's coherence. In a DP scheme, the decoherence is caused by the following propagator:
\begin{equation}
    \mathcal{L}_{{DP}} \hat{\sigma}(t) = - \frac{c^2}{2 \hbar^2} \int \mathrm{d}^3 x \ \mathrm{d}^3 y \ D^{tt}_{\mathcal{A}}(\vec{x}, \vec{y}) \left[ \hat{m}(\vec{x}), \left[ \hat{m}(\vec{y}) , \hat{\sigma}(t) \right] \right] \, .
\end{equation}
Now, to see whether there exists decoherence, we need to look at the off-diagonal elements of the reduced quantum state after applying the propagator, $\langle R_i | \mathcal{L}_{\text{DP}} \hat{\sigma}(t)| R_j \rangle \ \forall i \neq j$, which we do representatively by looking at
\begin{equation}\label{eq:off_diagonal_propagator}
    \langle R_1 | \mathcal{L}_{DP} \hat{\sigma}(t)| R_2 \rangle = \frac{c^2}{2 \hbar^2} \int \mathrm{d}^3 x \mathrm{d}^3 y D^{tt}_{\mathcal{A}}(\vec{x}, \vec{y}) \langle R_1 | [ \hat{m}(\vec{x})\hat{\sigma}(t)\hat{m}(\vec{y}) -\hat{m}(\vec{x})\hat{m}(\vec{y})\hat{\sigma}(t) + \hat{m}(\vec{y})\hat{\sigma}(t)\hat{m}(\vec{x}) - \hat{\sigma}(t)\hat{m}(\vec{y})\hat{m}(\vec{x}) ] | R_2 \rangle \, .
\end{equation}
Here, the reduced quantum state's form of Eq. \eqref{eq:superposition_state} can be inserted. Additionally, since the overlap of the rotational states is negligible, one can write (with the following notation not indicating a sum over $i$) $\langle R_i| \hat{m}(\vec{x}) | R_j \rangle = \langle \hat{m}(\vec{x})\rangle_i \delta_{ij}$, where we have defined $\langle \hat{m}(\vec{x})\rangle_i := \langle R_i| \hat{m}(\vec{x}) | R_i \rangle$. Through this expression we can also determine terms of the following form:
\begin{equation}
    \langle R_i | \hat{m}(\vec{x}) \hat{m}(\vec{y}) | R_j \rangle = \sum_k \langle R_i| \hat{m}(\vec{x}) | R_k \rangle \langle R_k| \hat{m}(\vec{y}) | R_j \rangle = \sum_k \langle\hat{m}(\vec{x})\rangle_i \delta_{ik} \langle \hat{m}(\vec{y})\rangle_k \delta_{kj} = \begin{cases}
        0, & i \neq j \, , \\ \langle\hat{m}(\vec{x})\rangle_i \langle\hat{m}(\vec{y})\rangle_j, & i=j \, .
    \end{cases}
\end{equation}
Hence, one can determine Eq. \eqref{eq:off_diagonal_propagator} to be the product of the respective differences between both rotational expectation values of the mass density:
\begin{equation}\label{eq:DP-decoherence}
    \langle R_1 | \mathcal{L}_{DP} \hat{\sigma}(t)| R_2 \rangle  = - \frac{c^2}{2 \hbar^2} \int \mathrm{d}^3 x \mathrm{d}^3 y D^{tt}_{\mathcal{A}}(\vec{x}, \vec{y}) \sigma^{12}(t) \left[\langle\hat{m}(\vec{x})\rangle_1 - \langle\hat{m}(\vec{x})\rangle_2 \right] \left[ \langle\hat{m}(\vec{y})\rangle_1 - \langle\hat{m}(\vec{y})\rangle_2 \right] \, .
\end{equation}
In our situation of interest, the considered object is symmetric about the rotation axis $\vec{n}$, implying $\hat{m}(\mathcal{R}^{-1}\vec{x}) = \hat{m}(\vec{x})$ for all such rotations $\mathcal{R}$. Therefore, applying the rotation operator to the mass density has no effect since $\hat{\mathcal{R}}^{\dagger} (\vec{\theta}) \hat{m}(\vec{x}) \hat{\mathcal{R}}(\vec{\theta}) = \hat{m}(\mathcal{R}^{-1}\vec{x}) = \hat{m}(\vec{x})$. Through that statement, one can deduce:
\begin{equation}
    \langle \hat{m}(\vec{x})\rangle_2 = \langle R_2 | \hat{m}(\vec{x}) | R_2 \rangle = \langle R_1 | \hat{\mathcal{R}}^{\dagger}(\vec{\theta}) \hat{m}(\vec{x}) \hat{\mathcal{R}}(\vec{\theta}) | R_1 \rangle = \langle R_1 | \hat{m}(\mathcal{R}^{-1}\vec{x}) | R_1 \rangle = \langle R_1 | \hat{m}(\vec{x}) | R_1 \rangle = \langle \hat{m}(\vec{x})\rangle_1 \, .
\end{equation}
Consequently, we find $\langle R_1 | \mathcal{L}_{DP} \hat{\sigma}(t)| R_2 \rangle = 0$, meaning that the rotation does not induce any decoherence in the standard collapse model.

The situation looks different in the post-Newtonian collapse model presented in this letter. Recall the rotational decoherence term of a rigid rotor:
\begin{equation}
    \mathcal{L}_{\text{rot}} \hat{\sigma}(t) = - \frac{1}{2\hbar^2} \int \mathrm{d}^3 x \ \mathrm{d}^3 y \ D^{kl}_{\mathcal{A}}(\vec{x}, \vec{y})\chi_{ki;lm}(\vec x,\vec y)\left[\hat{L}^i, \left[ \hat{L}^m, \hat{\sigma}(t) \right] \right] \, .
\end{equation}
We have $\chi_{ki;lm}(\vec x,\vec y)=m(\vec x\,)m(\vec y)\varepsilon_{kij}\varepsilon_{lmn}x^j y^n/I^2$ and denote $\varepsilon_{ijk}$ as the Levi-Civita symbol. Now, again, we look at the coherence of the reduced quantum state, this time under the rotational propagator:
\begin{equation}
    \langle R_1 | \mathcal{L}_{\text{rot}} \hat{\sigma}(t) | R_2 \rangle = - \frac{1}{2\hbar^2} \int \mathrm{d}^3 x \ \mathrm{d}^3 y \ D^{kl}_{\mathcal{A}}(\vec{x}, \vec{y})\chi_{ki;lm}(\vec x,\vec y)\langle R_1 |\left( \hat{L}^i \hat{L}^m \hat{\sigma}(t) - \hat{L}^i \hat{\sigma}(t) \hat{L}^m - \hat{L}^m \hat{\sigma}(t) \hat{L}^i + \hat{\sigma}(t) \hat{L}^m \hat{L}^i \right) | R_2 \rangle \, .
\end{equation}
By the same token as before, since the overlap between the rotational states is negligible, we have (with the following Kronecker delta notations never indicating a sum over $i$) $\langle R_i | \hat{L}^j |R_k \rangle = \langle \hat{L}^j \rangle_i \delta_{ik}$ where $ \langle\hat{L}^j\rangle_i := \langle R_i | \hat{L}^j | R_i \rangle$, and for combined terms with two angular momentum operators, we can define $\langle \hat{L}^j \hat{L}^k \rangle_i := \langle R_i | \hat{L}^j \hat{L}^k | R_i \rangle$ such that $\langle R_i | \hat{L}^j \hat{L}^k | R_l \rangle = \langle \hat{L}^j \hat{L}^k \rangle_i \delta_{il}$. These combined terms can be phrased as functions of the respective expectation values of the angular momentum operators as $\langle \hat{L}^j \hat{L}^k \rangle_i = \langle \hat{L}^j \rangle_i \langle  \hat{L}^k \rangle_i + \mathrm{Cov}_i(\hat{L}^j, \hat{L}^k)$, where $\mathrm{Cov}_i(\hat{L}^j, \hat{L}^k) = \langle (\hat{L}^j - \langle \hat{L}^j\rangle_i) (\hat{L}^k - \langle \hat{L}^k\rangle_i) \rangle_i$ denotes the covariance of both operators. 

When all terms are collected, the propagator of the coherence can be cast in a similar manner to the DP case. However, it now features expectation values and covariances of the angular momentum operator:
\begin{align}
    \langle R_1 | \mathcal{L}_{\text{rot}} \hat{\sigma}(t) | R_2 \rangle = - \frac{1}{2\hbar^2} \int \mathrm{d}^3 x \ \mathrm{d}^3 y \ D^{kl}_{\mathcal{A}}(\vec{x}, \vec{y})\chi_{ki;lm}(\vec x,\vec y) \sigma^{12}(t) \bigg[ \Big( \langle & \hat{L}^i\rangle_1 - \langle \hat{L}^i\rangle_2 \Big) \Big( \langle \hat{L}^m\rangle_1 - \langle \hat{L}^m\rangle_2 \Big) \notag \\ & + \mathrm{Cov}_1(\hat{L}^i, \hat{L}^m) + \mathrm{Cov}_2(\hat{L}^m, \hat{L}^i) \bigg] \, .
\end{align}
Apart from first moment terms resembling those of the DP decoherence rate of Eq. \eqref{eq:DP-decoherence}, we find nontrivial contributions from the respective covariances and variances as $\mathrm{Var}_i(\hat{L}^j) = \mathrm{Cov}_i(\hat{L}^j, \hat{L}^j) = \langle (\hat{L}^j)^2 \rangle_i - \langle \hat{L}^j \rangle_i^2$, which generically differ from each other and do not trivially vanish. 

Furthermore, in contrast to the DP case, even the first moments do not cancel when considering a rotation about a body's symmetry axis. To get an intuition behind this statement, we make use of the commutation relations of the angular momentum operator:
\begin{equation}
    \left[ \hat{L}^z, \hat{L}^x \right] = i \hbar \hat{L}^y , \quad \left[ \hat{L}^z, \hat{L}^y \right] = -i \hbar \hat{L}^x , \quad \left[ \hat{L}^z, \hat{L}^z \right] = 0 \, .
\end{equation}
Using these relations, we can apply the Baker-Campbell-Hausdorff formula to determine the action of the rotation operator on each angular momentum component $\hat{L}^i$. Under the assumption that the body rotates about its symmetry axis $\vec{n} = (0,0,1)^{\mathrm{T}}$, meaning that the rotation operator reads $\hat{\mathcal{R}}(\vec{\theta}) = \exp (- i \theta \hat{L}^z/\hbar)$, one obtains
\begin{equation}
    \hat{\mathcal{R}}^{\dagger}(\vec{\theta}) \hat{L}^x \hat{\mathcal{R}}(\vec{\theta}) = \hat{L}^x \cos \theta - \hat{L}^y \sin \theta \, ,
\end{equation}
\begin{equation}
    \hat{\mathcal{R}}^{\dagger}(\vec{\theta}) \hat{L}^y \hat{\mathcal{R}}(\vec{\theta}) = \hat{L}^y \cos \theta + \hat{L}^x \sin \theta \, ,
\end{equation}
\begin{equation}
    \hat{\mathcal{R}}^{\dagger}(\vec{\theta}) \hat{L}^z \hat{\mathcal{R}}(\vec{\theta}) = \hat{L}^z \, .
\end{equation}
For a theoretical background on commutation relations and the action of rotation operators, one may consult \cite{shankar2012principles}. From that we can see that the first moments of the decoherence term generally do not vanish. As an example, consider the $x$-component:
\begin{equation}
    \langle \hat{L}^x\rangle_2 = \langle R_2 | \hat{L}^x | R_2 \rangle = \langle R_1 | \hat{\mathcal{R}}^{\dagger} (\vec{\theta})\hat{L}^x \mathcal{R}(\vec{\theta}) | R_1 \rangle = \langle R_1 | \hat{L}^x | R_1 \rangle \cos \theta - \langle R_1 | \hat{L}^y | R_1 \rangle \sin \theta = \langle \hat{L}^x\rangle_1 \cos \theta - \langle \hat{L}^y\rangle_1 \sin \theta .
\end{equation}
It is clear that in general, $\langle \hat{L}^x\rangle_2 \neq \langle \hat{L}^x\rangle_1$, apart from special cases (for instance, if $\theta = 0$, or in finely tuned situations where trigonometric factors cancel). Therefore, we have found a simple class of scenarios (rotations around a symmetry axis for a body in rotational superposition) where the prediction of the standard DP scheme differs qualitatively from the effects that are ensued by a post-Newtonian description.

\section{A Detailed Derivation}\label{app:derivation}

This Appendix provides an extensive version of the hybrid dynamic's derivation presented in the main body. Again, we emphasize that we work along the lines of Di\'{o}si's derivation from Ref.~\cite{Diosi2011Hybrid}. Let us first introduce the four-component vector potential $A^{\mu}$ as classical degrees of freedom and the quantized mass current components $\hat{J}^{\mu}$ as quantum degrees of freedom. Thus, the phase space variables consist of $A^{\mu}$ as well as its conjugate momentum given by $ \vec{z} = \{ A^{\mu}(\vec{x}), \pi^{\mu}(\vec{x}) \} $. Their respective values are determined at certain points of a three-dimensional space denoted by $\vec{x} \in \mathbb{R}^3$. We define the quantum state and Hamiltonian as $\hat{\tau}(\vec{z},t)$ and $\hat{H}_Q(\vec{z},t)$, respectively. Furthermore, the isolated quantum system evolves according to the von Neumann equation
\begin{equation}
    \partial_t \hat{\tau}(\vec{z},t) = -\frac{i}{\hbar} [\hat{H}_Q(\vec{z},t),\hat{\tau}(\vec{z},t)] \, .    
\end{equation}
Similarly, we can introduce a classical Hamiltonian $H_C(\vec{z},t)$ and state $p(\vec{z},t)$ whose independent evolution (as described by the Liouville equation) relies on the Poisson bracket with respect to the phase space variables as 
\begin{equation}
    \partial_t p(\vec{z},t) = \{ H_C(\vec{z},t), p(\vec{z},t) \}_P \, .
\end{equation}
If one were to combine both states into a separable, hybrid state $ \hat{\rho}(\vec{z},t) = \hat{\tau}(\vec{z},t) p(\vec{z},t) $ with a composite Hamiltonian $ \hat{H}'(\vec{z},t) = \hat{H}_Q(\vec{z},t) + H_C(\vec{z},t) $, then the dynamics of this state would constitute a superposition of both sub-dynamics, given that the two states are independent. Thus, this entails 
\begin{equation}
    \partial_t \hat{\rho}(\vec{z},t) = -\frac{i}{\hbar} [\hat{H}'(\vec{z},t), \hat{\rho}(\vec{z},t) ] + \{ \hat{H}'(\vec{z},t), \hat{\rho}(\vec{z},t) \}_P \, .
\end{equation}
Now, by allowing the two systems to interact, a term $\hat{H}_{I}(\vec{z},t)$ must be added in the hybrid equation of motion such that the total hybrid Hamiltonian reads $\hat{H}(\vec{z},t) = \hat{H}_Q(\vec{z},t) + H_C(\vec{z},t) + \hat{H}_{I}(\vec{z},t)$. 

Here, it is important to acknowledge that the notation is hiding some of the underlying complexity. Rigorously speaking, one should define the hybrid state as a tensor product $\hat{\rho}(\vec{z},t) = \hat{\tau}(\vec{z},t) \otimes p(\vec{z},t) | \vec{z} \rangle \langle \vec{z} |$ of the quantum state with a diagonal basis in the phase space coordinates. In a similar fashion, the Hamiltonian must also be rewritten as $\hat{H}(\vec{z},t) = \hat{H}_Q(\vec{z},t) \otimes \mathbb{1} + \mathbb{1} \otimes H_C(\vec{z},t) | \vec{z} \rangle \langle \vec{z} | + \hat{H}_I(\vec{z},t) \otimes | \vec{z} \rangle \langle \vec{z} |$. However, to ensure the classicality of the system at all times, the evolution of the hybrid state has to maintain the block-diagonal structure. Therefore, the basis remains unaltered at every instant in time, thus allowing for the use of the simplified notation employed so far according to which the block-diagonal structure is kept implicit. At this point, using the hybrid Hamiltonian that includes the interaction term, one can state that the evolution of the hybrid state takes the form 
\begin{equation}
    \partial_t \hat{\rho}(\vec{z},t) = -\frac{i}{\hbar} [\hat{H}(\vec{z},t), \hat{\rho}(\vec{z},t) ] + \frac{1}{2} [ \{ \hat{H}(\vec{z},t), \hat{\rho}(\vec{z},t) \}_P  - \{ \hat{\rho}(\vec{z}, t), \hat{H}(\vec{z},t) \}_P  ] \, .
\end{equation}
This classical-quantum evolution is called Aleksandrov bracket and was first proposed in Ref.~\cite{Aleksandrov+1981+902+908}. The problem associated with this equation of motion is that it fails to ensure the positivity of the hybrid state, meaning that refinements to the interaction between classical and quantum systems are required. To this aim, one can assume that the interaction term coupling the classical gravitomagnetic vector potential to the quantized mass current, that is $\hat{H}_{I}(\vec{z}, t) = \int \mathrm{d}^3x \ \hat{J}^{\mu}(\vec{x}) \tilde A_{\mu}(\vec{x})$ (with $\tilde A^{\mu}=(\phi/c,\,4A^k)$), contains a classical noise term for both $\tilde A^{\mu}(\vec{x})$ and $\hat{J}^{\mu}(\vec{x})$. This noise enters the interaction Hamiltonian as $\hat{\mathcal{H}}_{I}(\vec{z},t) = \int \mathrm{d}^3 x \ \Big[ \hat{J}^{\mu}(\vec{x}) + \mathcal{J}^{\mu}(\vec{x}, t) \Big] \Big[ \tilde A_{\mu}(\vec{x}) + \mathcal{A}_{\mu}(\vec{x}, t) \Big]$. In so doing, a general noise Hamiltonian modifies the total Hamiltonian, which then becomes
\begin{align}\label{eq:general_noise_hamiltonian}
    \hat{\mathcal{H}}(\vec{z}, t) = \hat{H}_Q(\vec{z}, t) + H_C(\vec{z},t) + \hat{\mathcal{H}}_{I}(\vec{z},t) = \hat{H}(\vec{z},t) + \int \mathrm{d}^3 x \Big[ \hat{J}^{\mu}(\vec{x})\mathcal{A}_{\mu}(\vec{x},t) + \tilde A^{\mu}(\vec{x}) \mathcal{J}_{\mu}(\vec{x},t) + \mathcal{J}^{\mu}(\vec{x},t) \mathcal{A}_{\mu}(\vec{x},t) \Big] \, .
\end{align}
As customary in these approaches to collapse models, the noise is taken to be Gaussian and white, so that their first moments (\emph{i.e.}, their respective noise averages) $\langle \bullet \rangle$ vanish, whereas the second moments are component-wise spatially-dependent coefficients sharply peaked in time, namely
\begin{align}
        \langle \mathcal{J}^{\mu}(\vec{x}, t) \rangle = \langle \mathcal{A}^{\mu}(\vec{x}, t) \rangle = 0 \, , \quad \langle \mathcal{J}^{\mu}(\vec{x},t) \mathcal{J}^{\nu}(\vec{y},t') \rangle = D^{\mu \nu}_{\mathcal{J}}(\vec{x},\vec{y}) \delta(t-t') \, , \quad \langle \mathcal{A}^{\mu}(\vec{x},t) \mathcal{A}^{\nu}(\vec{y},t') \rangle = D^{\mu \nu}_{\mathcal{A}}(\vec{x},\vec{y}) \delta(t-t') \, .
\end{align}
Now, to ensure the positivity of the hybrid state in Eq.~\eqref{aleks}, one can refine the dynamics of the hybrid state by averaging over the noise, thereby remaining with
\begin{equation}
    \partial_t \hat{\rho}(\vec{z},t) = \left\langle - \frac{i}{\hbar} [\hat{\mathcal{H}}(\vec{z},t), \hat{\rho}(\vec{z},t) ] + \frac{1}{2} \Big[ \{ \hat{\mathcal{H}}(\vec{z},t), \hat{\rho}(\vec{z},t) \}_P - \{ \hat{\rho}(\vec{z}, t), \hat{\mathcal{H}}(\vec{z},t) \}_P  \Big] \right\rangle \, .
\end{equation}
For the noise Hamiltonian of Eq. \eqref{eq:general_noise_hamiltonian}, we can then define the quantum noise Hamiltonian $\hat{\mathcal{H}}_{Q}(\vec{z},t) := \int \mathrm{d}^3 x \ \hat{J}^{\mu}(\vec{x}) \mathcal{A}_{\mu}(\vec{x},t)$, the classical noise Hamiltonian $\mathcal{H}_{C}(\vec{z},t) := \int \mathrm{d}^3 x \ \tilde A^{\mu}(\vec{x}) \mathcal{J}_{\mu}(\vec{x},t)$, and additionally assume $\langle \mathcal{J}^{\mu}(\vec{x},t)\mathcal{A}^{\nu}(\vec{y},t') \rangle = 0$. To see how both noise Hamiltonians alter the dynamics, one can look at expansions of the time-dependent state. Starting from a state that is subject to the quantum noise Hamiltonian $\hat{\mathcal{H}}_{Q}(\vec{z},t)$ only, one can write its time evolution from an initial state at $t_0$ as
\begin{equation}
    \hat{\rho}(\vec{z},t) = \mathcal{T} \exp{-i \int_{t_0}^t \mathrm{d}\tau \ \hat{\mathcal{H}}_{Q}(\vec{z}, \tau)/\hbar} \hat{\rho}(\vec{z},t_0) \Tilde{\mathcal{T}} \exp{i \int_{t_0}^t \mathrm{d}\tau \ \hat{\mathcal{H}}_{Q}(\vec{z},\tau)/\hbar} \, ,
\end{equation}
where $\mathcal{T}$ denotes the time-ordering operator and $\Tilde{\mathcal{T}}$ the anti-time-ordering operator. Expanding the time-evolution operators in a Dyson series up to second order yields the following expression:
\begin{align}
    \hat{\rho}(\vec{z},t) \approx \hat{\rho}(\vec{z},t_0) - \frac{i}{\hbar} \int_{t_0}^t \mathrm{d}\tau \ \left[ \hat{\mathcal{H}}_{Q}(\vec{z},\tau), \hat{\rho}(\vec{z},t_0) \right] - \frac{1}{2\hbar^2} \int_{t_0}^t \mathrm{d}\tau \ \mathrm{d}\tau' \ \left[ \hat{\mathcal{H}}_{Q}(\vec{z},\tau), \left[\hat{\mathcal{H}}_{Q}(\vec{z},\tau') , \hat{\rho}(\vec{z},t_0)\right] \right] \, .
\end{align}
By taking the noise average and inserting the definition of the quantum noise Hamiltonian $\hat{\mathcal{H}}_{Q}(\vec{z},t)$, we see that the first order contribution vanishes, whereas the zeroth and second order survive and give rise to
\begin{align}
    \frac{\hat{\rho}(\vec{z},t) - \hat{\rho}(\vec{z},t_0)}{t-t_0} = - \frac{1}{2\hbar^2} \int \mathrm{d}^3 x \ \mathrm{d}^3 y \ D_{\mathcal{A}}^{\mu\nu} (\vec{x}, \vec{y}) \left[ \hat{J}_{\mu}(\vec{x}), \left[ \hat{J}_{\nu}(\vec{y}) , \hat{\rho}(\vec{z},t_0) \right] \right] \, .
\end{align}
Clearly, the left-hand side of the above equation can be turned into a time derivative upon substituting $\Delta t = t-t_0$ and taking the limit $\Delta t \rightarrow 0$. Then, one can replace $t_0 \rightarrow t$ to reach the final form of the dynamical equation produced by the quantum noise Hamiltonian $\hat{\mathcal{H}}_{Q}(\vec{z},t)$, that is
\begin{align}
    \partial_t \hat{\rho}(\vec{z},t) \approx - \frac{1}{2\hbar^2} \int  \mathrm{d}^3 x \ \mathrm{d}^3 y \ D_{\mathcal{A}}^{\mu\nu} (\vec{x}, \vec{y}) \left[ \hat{J}_{\mu}(\vec{x}), \left[ \hat{J}_{\nu}(\vec{y}) , \hat{\rho}(\vec{z},t) \right] \right] \, .
\end{align}
Hence, the noise-averaged evolution contains a double-commutator term that accounts for the noise from the four-component vector potential fluctuations on the quantum matter. The same procedure can be applied to the noise that the classical system is subjected to by expanding the initial state evolution 
\begin{equation}
\hat{\rho}(\vec{z},t) = \mathcal{T} \exp{\int_{t_0}^t \mathrm{d} \tau \ \left\{ \mathcal{H}_C(\vec{z},\tau), \bullet \right\}_P} \hat{\rho}(\vec{z},t_0) \, ,
\end{equation}
which is generated by the Poisson bracket. Again, the expansion is carried out to second order (since the first order vanishes in the noise average), and its form
\begin{align}
    \hat{\rho}(\vec{z},t) \approx \hat{\rho}(\vec{z},t_0) + \int_{t_0}^t \mathrm{d} \tau \ \left\{ \mathcal{H}_C(\vec{z},\tau), \hat{\rho}(\vec{z},t_0) \right\}_P + \frac{1}{2} \int_{t_0}^t \mathrm{d} \tau \ \mathrm{d} \tau' \ \left\{ \mathcal{H}_C(\vec{z},\tau), \left\{ \mathcal{H}_C(\vec{z},\tau'), \hat{\rho}(\vec{z},t_0) \right\}_P \right\}_P
\end{align}
contains a nested Poisson bracket. By constructing the time derivative on the l.h.s. and by averaging over the noise, the dynamical evolution becomes
\begin{align}
    \partial_t \hat{\rho}(\vec{z},t) \approx \frac{1}{2} \int \mathrm{d}^3 x \ \mathrm{d}^3 y  \ D^{\mu \nu}_{\mathcal{J}}(\vec{x}, \vec{y}) \left\{\tilde A_{\mu}(\vec{x}), \left\{ \tilde A_{\nu}(\vec{y}), \hat{\rho}(\vec{z},t) \right\}_P \right\}_P \, .
\end{align}
Thus, the entire averaged evolution of a hybrid state accounts for the noise of both systems on one another. Interestingly, this formalism constitutes a natural relativistic extension to the dynamics of Ref. \cite{Diosi2011Hybrid}. This can be seen by recasting the interaction Hamiltonian to make explicit the hybrid dynamics introduced in Ref. \cite{Diosi2011Hybrid} for the quantized mass density $\hat{m}(\vec{x})$ and the gravitational potential $\Phi(\vec{x})$ (see \eqref{eq:interaction_definition}). If all distinct terms are written out explicitly for $k,l \in \{ 1,2,3 \}$, the hybrid, noise-averaged dynamics reads:
\begin{align}\label{eq:big_general_dynamics}
    \partial_t \hat{\rho}(\vec{z},t) & = - \frac{i}{\hbar} [\hat{H}(\vec{z},t), \hat{\rho}(\vec{z},t) ] + \frac{1}{2} \Big[ \{ \hat{H}(\vec{z},t), \hat{\rho}(\vec{z},t) \}_P - \{ \hat{\rho}(\vec{z}, t), \hat{H}(\vec{z},t) \}_P  \Big] \notag \\ & - \frac{c^2}{2\hbar^2} \int \mathrm{d}^3 x \ \mathrm{d^3}y \ D^{tt}_{\mathcal{A}}(\vec{x},\vec{y}) \left[ \hat{m}(\vec{x}),\left[ \hat{m}(\vec{y}) , \hat{\rho}(\vec{z}, t)\right] \right] + \frac{1}{2c^2} \int \mathrm{d}^3 x \ \mathrm{d^3}y \ D^{tt}_{\mathcal{J}}(\vec{x},\vec{y}) \left\{ \Phi(\vec{x}), \left\{ \Phi(\vec{y}), \hat{\rho}(\vec{z},t) \right\}_P \right\}_P \notag \\ & - \frac{c}{2\hbar^2} \int \mathrm{d}^3 x \ \mathrm{d^3}y \ D^{tl}_{\mathcal{A}}(\vec{x},\vec{y}) \left[ \hat{m}(\vec{x}),\left[ \hat{J}_{l}(\vec{y}) , \hat{\rho}(\vec{z}, t)\right] \right] + \frac{1}{2c}  \int \mathrm{d}^3 x \ \mathrm{d^3}y \ D^{tl}_{\mathcal{J}}(\vec{x},\vec{y}) \left\{ \Phi(\vec{x}), \left\{ \tilde A_l(\vec{y}), \hat{\rho}(\vec{z},t) \right\}_P \right\}_P \notag \\ & - \frac{c}{2\hbar^2} \int \mathrm{d}^3 x \ \mathrm{d^3}y \ D^{kt}_{\mathcal{A}}(\vec{x},\vec{y}) \left[ \hat{J}_{k}(\vec{x}),\left[ \hat{m}(\vec{y}) , \hat{\rho}(\vec{z}, t)\right] \right] + \frac{1}{2c}  \int \mathrm{d}^3 x \ \mathrm{d^3}y \ D^{kt}_{\mathcal{J}}(\vec{x},\vec{y}) \left\{ \tilde A_k(\vec{x}), \left\{ \Phi(\vec{y}), \hat{\rho}(\vec{z},t) \right\}_P \right\}_P \notag \\ & - \frac{1}{2\hbar^2} \int \mathrm{d}^3 x \ \mathrm{d^3}y \ D^{kl}_{\mathcal{A}}(\vec{x},\vec{y}) \left[ \hat{J}_{k}(\vec{x}),\left[ \hat{J}_{l}(\vec{y}) , \hat{\rho}(\vec{z}, t)\right] \right] + \frac{1}{2}  \int \mathrm{d}^3 x \ \mathrm{d^3}y \ D^{kl}_{\mathcal{J}}(\vec{x},\vec{y}) \left\{ \tilde A_k(\vec{x}), \left\{ \tilde A_l(\vec{y}), \hat{\rho}(\vec{z},t) \right\}_P \right\}_P \, .
\end{align}

\section{The Necessity of Fundamental Noise}\label{app:noise_necessity}
\begin{figure*}
    \centering
    \begin{subfigure}{0.49\linewidth}
        \centering
        \includegraphics[width=\linewidth]{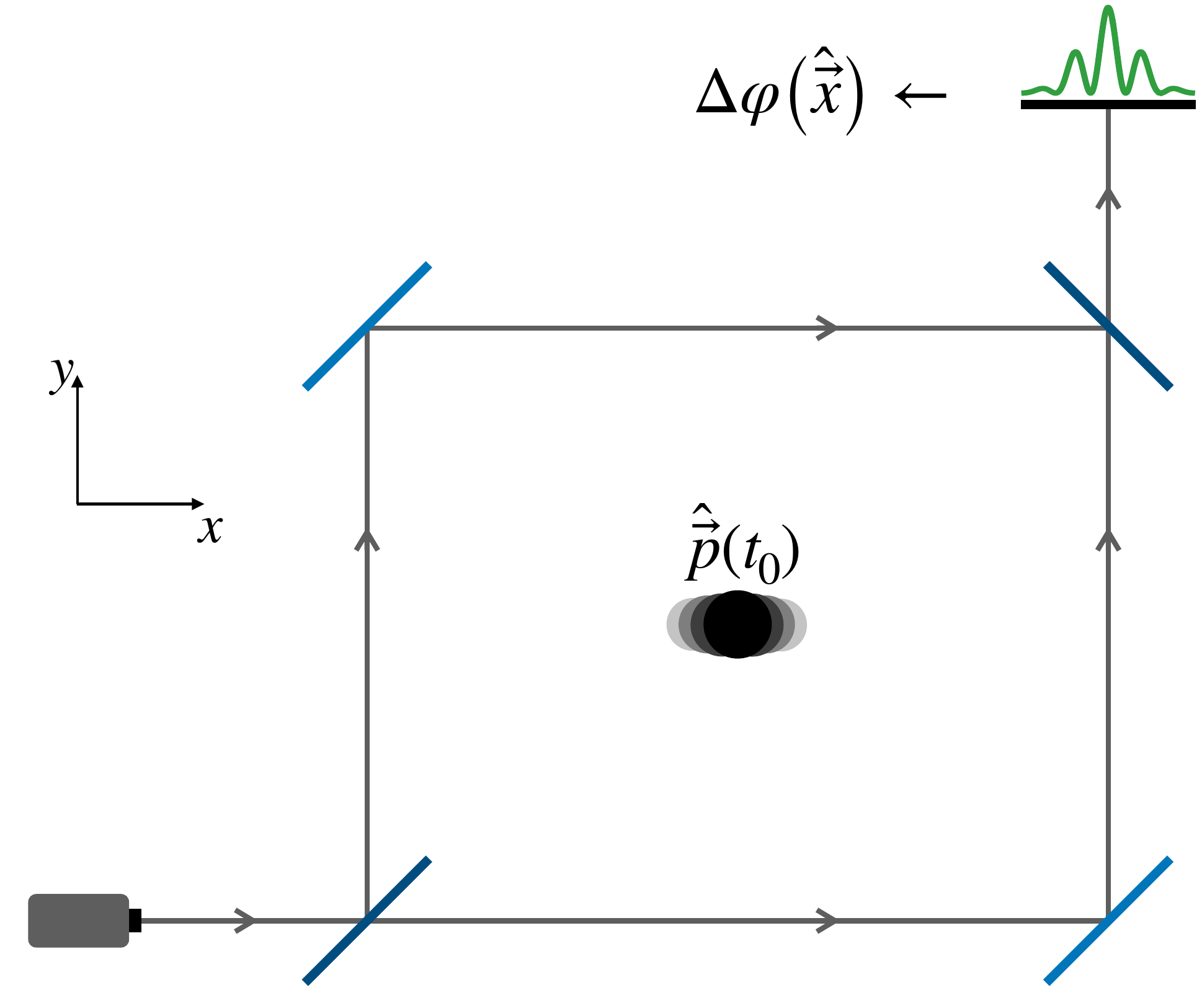}
        \caption{}
        \label{fig:xpinterferometer}
    \end{subfigure}
    \hfill
    \begin{subfigure}{0.49\linewidth}
        \centering
        \includegraphics[width=\linewidth]{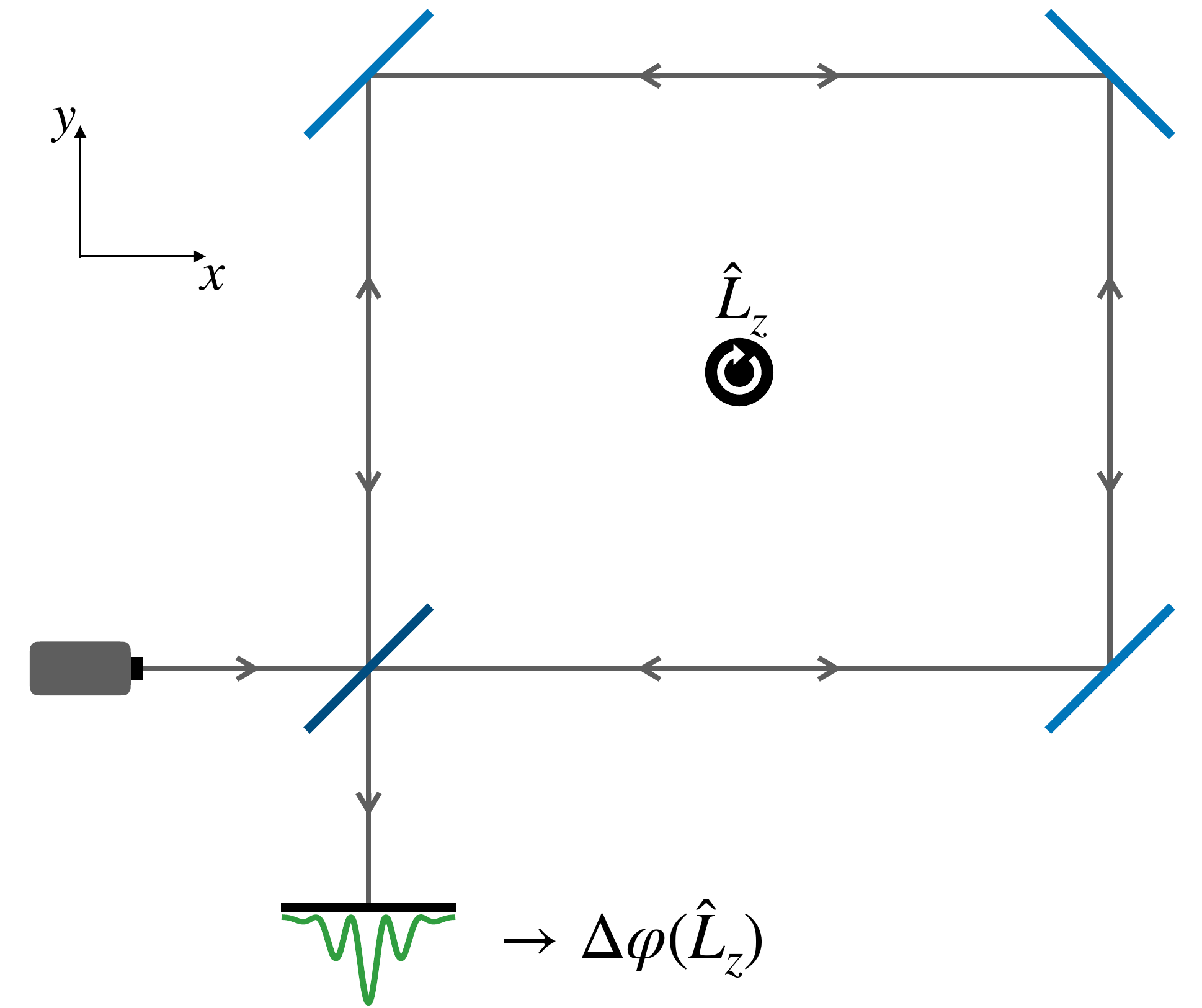}
        \caption{}
        \label{fig:Sagnac}
    \end{subfigure}
    
    \caption{(a) Sketch of a quantum system prepared in a momentum eigenstate with $\hat{\vec{p}}(t_0)$. It sources a gravitational field which can be read out from the phase shift $\Delta \varphi \propto \int \mathrm{d}t \ 1/|\hat{\vec{x}}|$. Since the phase can be obtained without performing a measurement on the system, the uncertainty relation \eqref{eq:xp_uncertainty} is violated.
    (b) Sketch of a rotating quantum system enclosed by a ring interferometer. In the displayed setup, the interferometer plane corresponds to the $x$-$y$ plane. The classical field $\vec{A}$ is sourced by $\hat{L}_z$, thus leading to a phase shift $\Delta \varphi \propto \hat{L}_z$. If additional interferometers are placed at the $x$-$z$ and $y$-$z$ planes, all angular momentum components could be measured with arbitrarily high precision, which violates the uncertainty relations of Eq. ~\eqref{eq:uncertainty_relation}.}
    
    \label{fig:combined_interferometers}
\end{figure*}

Fundamental classical noise on the mass density is a consistency requirement for typical classical-quantum hybrid dynamics~\cite{diosi1995quantum, Diosi2011Hybrid, oppenheim2023postquantum}. When dealing with a formalism that postulates such a noise for the rotational degrees of freedom, it is apt to consider a Gedankenexperiment showing the necessity of doing so (sharing similarities with the reasoning of \cite{oppenheim2023postquantum}). Quantum physics requires the following to hold for the variances of the angular momentum operator $\hat{\vec{L}}$ \cite{shankar2012principles}:
\begin{equation}\label{eq:uncertainty_relation}
    \sigma_{\hat{L}_i} \sigma_{\hat{L}_j} \geq \frac{\hbar}{2} |\langle \hat{L}_k \rangle|, \qquad \forall \ i \neq j \neq k \neq i \, .
\end{equation}
Likewise, the Heisenberg uncertainty relation restricting the variances of a quantum system's position and momentum operator reads:
\begin{equation}\label{eq:xp_uncertainty}
    \sigma_{\hat{x}} \sigma_{\hat{p}} \geq \frac{\hbar}{2} \, .
\end{equation}
Let us now consider a massive quantum system prepared in a sharp momentum eigenstate of $\hat{\vec{p}}(t_0)$. Usually, if one aims to determine the position operator $\hat{\vec{x}}$, the precise knowledge of the momentum eigenstate would be lost as a result of the measurement. However, in a noiseless hybrid theory of gravity, the quantum state sources a purely classical, deterministic gravitational field $\Phi(\vec x)$ whose value depends on the system's position. If said system is now enclosed by an interferometer (as depicted in Fig.~\ref{fig:xpinterferometer}), the phase difference $\Delta \varphi$ that can be inferred from the interference pattern is sensitive to the potential along the paths, such that $\Delta \varphi \propto \int \mathrm{d}t \, \Phi(\hat{\vec{x}}) \propto \int \mathrm{d}t/|\hat{\vec{x}}|$, where the second step is valid in the far-field approximation (\emph{i.e.}, when the extent of the quantum system is much smaller than the extent of the interferometer). In the ideal noiseless limit, this would allow for the inference of the system's position up to arbitrary precision from $\Delta \varphi$ without a corresponding back-action channel on the quantum state. Under this assumption, the information on the exact momentum eigenstate would not be degraded while the position is measured to arbitrary accuracy, thereby implying a violation of the uncertainty principle~\eqref{eq:xp_uncertainty}.

A similar argument can be made using the angular momentum uncertainty relation \eqref{eq:uncertainty_relation}. In particular, if the gravitomagnetic potential $\vec{A}$ is treated as being a classical object without any noisy feature, the full information about the quantum observable $\hat{\vec{L}}$ leaks into classical spacetime. To show this, let us now consider a ring (\emph{i.e.}, a Sagnac \cite{sagnac1913ether}) interferometer that encloses a rotating quantum system (as depicted in Fig. \ref{fig:Sagnac}). When light is sent through the arms of the interferometer, it picks up an effective phase $\Delta \varphi$ that depends on the gravitomagnetic potential integrated along the light-path as $\Delta \varphi = \int_{\mathcal{C}} \mathrm{d}\vec{l} \ \cdot \vec{A}$. Invoking Stokes' theorem, this can be cast as an area integral in terms of the gravitomagnetic field $\vec{B} = \nabla \wedge \vec{A} $ from the quantized version of \eqref{eq:magneticfield} and the interferometer plane's normal vector $\vec{n}$, that is $\Delta \varphi = \int_{\mathcal{S}} \mathrm{d}\mathcal{S} \  \vec{n} \cdot \vec{B}$. Consequently, since $\vec{B} \propto \hat{\vec{L}}$ in the far-field limit (i.e. when the extent of the object is negligible with respect to the interferometer), the phase can be used to read out angular momentum components up to arbitrary precision without disturbing the quantum system, since we assumed the gravitomagnetic potential to be classical and deterministic. Thus, this implies $\Delta \varphi \propto \vec{n} \cdot \hat{\vec{L}}$. Therefore, when three interferometers are arranged around the quantum system in a way that their normal vectors form a basis for $\mathbb{R}^3$ (e.g., $\vec{n}_k = \vec{e}_k$ such that $\Delta \varphi \propto \vec{n}_k \cdot \hat{\vec{L}} = \hat{L}_k$), all components of $\hat{\vec{L}}$ can be read out from the classical field up to an arbitrary precision, thereby violating the uncertainty principle for the angular momentum operator~\eqref{eq:uncertainty_relation}.
\section{Reduced Quantum State Dynamics}\label{app:reduced_dynamics}
From the general dynamics of Eq. \eqref{eq:big_general_dynamics}, a Master equation for the reduced quantum state can be obtained by integrating over the classical degrees of freedom. The reduced quantum state is defined as
\begin{equation}\label{eq:definition_reduced_quantum_state}
    \hat{\sigma}(t) := \int \mathrm{d}^8 z \ \hat{\rho}(\vec{z}, t) = \int \mathcal{D} \Tilde{A} \ \mathcal{D} \Tilde{\pi} \ \hat{\rho}[\Tilde{A}^{\mu}, \Tilde{\pi}^{\mu}, t] \, .
\end{equation}
In this notation, $\Tilde{A}^{\mu}$ and its conjugate momentum $\Tilde{\pi}^{\mu}$ make up the classical degrees of freedom $\vec{z}$. This is the reason why functional analysis must be considered henceforth, with $\int\mathcal{D}\tilde A$ expressing an integration over all the allowed field configurations. 

Now, as we are interested in the decoherence/diffusion terms of the reduced quantum state's dynamics, we denote the non-Hamiltonian part of the reduced quantum systems dynamics as
\begin{align}
    \partial_t \hat{\sigma}_{\backslash \hat{H}}(t) = & - \frac{1}{2\hbar^2} \int \mathrm{d}^3 x \ \mathrm{d^3}y \int \mathcal{D} \Tilde{A} \ \mathcal{D} \Tilde{\pi} \ D^{\mu \nu}_{\mathcal{A}}(\vec{x},\vec{y}) \left[ \hat{J}_{\mu}(\vec{x}),\left[ \hat{J}_{\nu}(\vec{y}) , \hat{\rho}(\vec{z}, t)\right] \right] \notag \\[2mm] & + \frac{1}{2} \int \mathrm{d}^3 x \ \mathrm{d^3}y \int \mathcal{D} \Tilde{A} \ \mathcal{D} \Tilde{\pi} \ D^{\mu \nu}_{\mathcal{J}}(\vec{x},\vec{y}) \left\{ \tilde A_{\mu}(\vec{x}), \left\{ \tilde A_{\nu}(\vec{y}), \hat{\rho}(\vec{z},t) \right\}_P \right\}_P  \, .
\end{align}
In the double-commutator term, the only object which depends on the classical degrees of freedom $\vec{z}$ is the hybrid state $\hat{\rho}(\vec{z}, t)$, meaning that Eq. \eqref{eq:definition_reduced_quantum_state} can be directly applied. For the nested Poisson bracket term, we can use the definition of the functional Poisson bracket to obtain
\begin{equation}
    \left\{ \Tilde{A}_{\nu}(\vec{y}), \hat{\rho} \right\}_P = \int \mathrm{d}^3x\left(\underbrace{\frac{\delta\Tilde{A}_{\nu}(\vec{y})}{\delta\Tilde{A}_{\mu}(\vec{x})}}_{= \delta^{\mu}_{\nu}\delta(\vec x-\vec y)} \frac{\delta\hat{\rho}(\vec{z},t)}{\delta\Tilde{\pi}^{\mu}(\vec{x})} - \underbrace{\frac{\delta \Tilde{A}_{\nu}(\vec{y})}{\delta \Tilde{\pi}_{\mu}(\vec{x})}}_{=0} \frac{\delta \hat{\rho}(\vec{z},t)}{\delta \Tilde{A}^{\mu}(\vec{y})}\right) = \delta^{\mu}_{\nu} \frac{\delta\hat{\rho}(\vec{z},t)}{\delta\Tilde{\pi}^{\mu}(\vec{y})} = \frac{\delta \hat{\rho}(\vec{z},t)}{\delta \Tilde{\pi}^{\nu}(\vec{y})} \, ,
\end{equation}
where $\delta^{\mu}_{\nu}$ is the Kronecker delta. This conveniently enables one to also find an expression for the nested Poisson bracket, which reads
\begin{equation}
    \left\{ \tilde A_{\mu}(\vec{x}), \left\{ \tilde A_{\nu}(\vec{y}), \hat{\rho}(\vec{z}, t)\right\}_P \right\}_P = \frac{\delta^2 \hat{\rho}(\vec{z}, t)}{\delta \Tilde{\pi}^{\mu}(\vec{x}) \delta \Tilde{\pi}^{\nu}(\vec{y})} \, .
\end{equation}
Plugging this relation into the reduced quantum system's dynamics yields the following expression for the nested Poisson bracket term:
\begin{equation}
    \mathcal{I} = \frac{1}{2} \int \mathrm{d}^3 x \ \mathrm{d}^3 y \int \mathcal{D} \Tilde{A} \ \mathcal{D} \Tilde{\pi} \ D^{\mu \nu}_{\mathcal{J}}(\vec{x}, \vec{y}) \frac{\delta^2 \hat{\rho}(\vec{z}, t)}{\delta \Tilde{\pi}^{\mu}(\vec{x}) \delta \Tilde{\pi}^{\nu}(\vec{y})} \, .
\end{equation}
We can now separately look at two cases. The first case is the part of the integral that goes over $\Tilde{\pi}_{\nu}(\vec{y})$, where we have derivatives of different indices, i.e. $\mu \neq \nu$. The second case is considering the situation where $\mu = \nu$. For the first case, $\mathcal{I}$ contains integrals of the form
\begin{equation}
    \int_{-\infty}^{\infty} \mathcal{D}\Tilde{\pi} \ \frac{\delta \hat{\rho}(\vec{z}, t)}{\delta \Tilde{\pi}^{\nu}(\vec{y})} = \lim_{\Tilde{\pi}_{\nu}(\vec{y}) \rightarrow \infty} \hat{\rho}(\vec{z}, t) - \lim_{\Tilde{\pi}_{\nu}(\vec{y}) \rightarrow - \infty} \hat{\rho}(\vec{z}, t) \, .
\end{equation}
The second case leads to a similar integration, with the only difference being that second order derivatives appear inside the integrand:
\begin{equation}
    \int_{-\infty}^{\infty} \mathcal{D}\Tilde{\pi} \ \frac{\delta^2 \hat{\rho}(\vec{z},t)}{\delta \Tilde{\pi}_{\nu}(\vec{x}) \delta \Tilde{\pi}_{\nu}(\vec{y})} = \lim_{\Tilde{\pi}_{\nu}(\vec{y}) \rightarrow \infty} \frac{\delta \hat{\rho}(\vec{z}, t)}{\delta \Tilde{\pi}_{\nu}(\vec{x})} - \lim_{\Tilde{\pi}_{\nu}(\vec{y}) \rightarrow - \infty} \frac{\delta \hat{\rho}(\vec{z}, t)}{\delta \Tilde{\pi}_{\nu}(\vec{x})} \, .
\end{equation}
Since the hybrid state must be well-behaved (i.e. normalizable), it must vanish whenever the absolute value of at least one component of the classical degrees of freedom approaches infinity. Moreover, the same logic applies for the hybrid state's derivative with respect to any component of the classical degrees of freedom, in particular. Thus one can write:
\begin{equation}
    \lim_{\Tilde{\pi}_{\nu}(\vec{y})\rightarrow \pm \infty} \hat{\rho}(\vec{z}, t) = \lim_{\Tilde{\pi}_{\nu}(\vec{y}) \rightarrow \pm \infty} \frac{\delta\hat{\rho}(\vec{z}, t)}{\delta \Tilde{\pi}_{\nu}(\vec{x})} = 0 \, , \quad \forall \nu \, .
\end{equation}
Therefore, the nested Poisson bracket term vanishes as $\mathcal{I} = 0$, such that the non-Hamiltonian part of the reduced quantum state's dynamics solely consists of the double commutator term
\begin{equation}
    \partial_t \hat{\sigma}_{\backslash \hat{H}}(t) = - \frac{1}{2\hbar^2} \int \mathrm{d}^3 x \ \mathrm{d^3}y \ D^{\mu \nu}_{\mathcal{A}}(\vec{x},\vec{y}) \left[ \hat{J}_{\mu}(\vec{x}),\left[ \hat{J}_{\nu}(\vec{y}) , \hat{\sigma}(t)\right] \right] \, .
\end{equation}

\end{document}